# Effects of Longitudinal Spin Current Density Gradient on Spin-Orbit Torque Switching of Perpendicular Magnetization

Guowen Gong,[1,2] Qianbiao Liu,[1] Lijun Zhu[1,2,*]

[1]*State Key Laboratory of Semiconductor Physics and Chip Technologies, Institute of Semiconductors, Chinese Academy of Sciences, Beijing 100083, China*

[2]*Center of Materials Science and Optoelectronics Engineering, University of Chinese Academy of Sciences, Beijing 100049, China*

**ljzhu@semi.ac.cn*

It has remained a critical open question as to whether a longitudinal spin current gradient (e.g., due to a gradient in the thickness, composition, or width) can replace the longitudinal magnetic field required for deterministic spin-orbit torque switching of normal metal/ferromagnet heterostructures with perpendicular magnetic anisotropy. Here, we report robust micromagnetic and experimental evidence that any realistic longitudinal spin-current density gradient cannot be an effective replacement for a longitudinal magnetic field to enable deterministic switching of a perpendicular magnetic anisotropy device. Instead, the longitudinal spin current density can only modify the nucleation and pulse-timing-sensitive oscillations of magnetic domains and thus the picosecond-scale strip-like indeterministic switching windows. The same conclusions hold robustly when the transverse effective field associated with the spin-current pulse (the sum of the Oersted field and field-like torque) is taken into account. The indeterministic switching prevents applications in the presence of finite device-to-device variations and drifts in write current pulse and thermal fluctuations of device parameters. We also experimentally show that, instead of a longitudinal spin current density gradient, perpendicular spins are much more effective in deterministic switching of perpendicular spin torque devices.

Deterministic electrical reversal of nanoscale magnetization is a central problem in spintronics. In spin-orbit-torque (SOT) devices, an in-plane charge current in a spin Hall metal generates a transversely polarized spin current that exerts SOTs on the neighboring ferromagnet [1-4]. However, the transverse spin polarizations cannot deterministically switch a magnetization with perpendicular magnetic anisotropy (PMA) due to their orthogonal orientations. In such orthogonal spin torque devices, increasing the pulse duration or pulse amplitude by a moderate amount beyond the critical switching value over-switches the magnetic layer back to its initial state (see schematics in Fig. 1(a))[5,6]. Consequently, a longitudinal magnetic field ($H_x$)[2,7], or equivalent effective field [8-13], parallel to the charge current is typically required to assist the switching of perpendicular magnetization by transverse spins. However, deterministic all-electrical switching of magnetization is indispensable for practical memory and logic applications. As schematically shown in Figs. 1(b) and 1(c), for deterministic all-electrical switching, the device is reliably switched as soon as the amplitude of the write current pulse exceeds its threshold at a given pulse width. Only devices with such robust deterministic switching could remain reliable against the drifts of material parameters (due to thermal fluctuations and device-to-device variations) and write current pulse profile (amplitude, width, rise/drop edges, and clocking).

Some early experiments suggested partial switching of PMA normal metal/ferromagnet (NM/FM) heterostructures by a current flow (Fig. 1(c)) *in the presence of* a longitudinal gradient in the width [14], the NM thickness, the insertion layer thickness [15], or the NM composition [16] (no detailed characteristics of the device geometries were provided via optical or electron microscopy imaging). It was speculated that the *longitudinal* structural gradient induces a longitudinal spin current density gradient ($\frac{\partial j_s}{\partial x}$) via the charge current density gradient ($\nabla_J$) and that the spin current density gradient functioned as a longitudinal effective magnetic field ($H_x \propto \nabla_J \propto \frac{\partial j_s}{\partial x}$) or an effective torque of $-M \times \nabla_J$ [14-17]. So far, there has been no supporting modeling or micromagnetic verification of the effects of the spin current density gradient, leaving the effectiveness of the field-free partial switching mechanism in these systems an open question. Experimentally, it is also challenging to disentangle the dominant effect that enabled the all-electrical switching. Even the role of the charge current gradient is not experimentally verifiable since a longitudinal composition or thickness gradient may vary the current density as well as many other critical thickness/composition-sensitive parameters, including the magnetic anisotropy [18], the interfacial spin transparency [19], the Dzyaloshinskii–Moriya Interaction (DMI)[20-22], the spin Hall ratio [4,23], the Oersted field, and the damping.

Here, we report robust micromagnetic and experimental evidence that any realistic longitudinal spin current gradient is insufficient to promote deterministic SOT switching of perpendicular magnetization. We show that any realistic spin-current density gradient, together with the associated transverse effective field (Oersted field and field-like torque), is not an effective replacement for a longitudinal magnetic field to yield deterministic SOT switching of a PMA device.

*Micromagnetic model*. As shown in Fig.1c, there are multiple ways to introduce a *longitudinal* gradient in the charge current density and thus spin current density (in the $x$ direction) in a HM/FM bilayer, e.g., thickness wedge, composition wedge, and width wedge. To verify the effects of such longitudinal spin and charge current gradients, we simulate a NM/FM bilayer by taking into account the damping-like torque associated with the spin current and the effective transverse field from the current-induced Oersted field and the fieldlike torque. As illustrated in Fig. 1(d), we consider a pulsed charge current density of $J_c = J_{c,0} + \nabla_J x$ within the HM layer, where $J_{c,0}$ is the average current density. Meanwhile, the charge current density will generate a pulsed transverse field ($H_y$) with a longitudinal gradient via the Oersted field and the field-like torque in the FM layer, i.e., $H_y = H_{y,0} + \nabla_H x$, with $H_{y,0}$ and $\nabla_H$ being the average value and the gradient of the transverse effective field associated with the longitudinal charge current. The FM layer is a circular cylinder of 100 nm in diameter and 1 nm in thickness. Other parameters include the exchange stiffness ($A$) of 20 pJ/m, the DMI constant ($D$) of 1 mJ/m$^2$, the uniaxial perpendicular anisotropy ($K_u$) of 0.8×10$^6$ J/m$^3$, the saturation magnetization ($M_s$) of 1.1×10$^6$ A/m, and magnetic damping constant ($\alpha$) of 0.1. The damping-like SOT efficiency of the HM/FM bilayer is fixed at 0.2, which is within the range of typical strong spin Hall metal/FM systems [24].

*Switching in the absence of current density gradient*. We first stimulate the switching in the uniform current limit ($\nabla_J = \nabla_H \equiv 0$). As shown in Fig. 2a,b, the device exhibits deterministic switching by a current (5 ns in duration) only in the presence of a significant assisting longitudinal magnetic field ($H_x$), while a large transverse magnetic field ($H_y$) does not help. In contrast, the device shows some indeterministic switching events in the absence of $H_x$, regardless of the magnitude of $H_y$. This observation coincides well with the experiments [3]. Due to the orthogonal orientations of the perpendicular magnetization and the transverse spins, the transverse spins do not have any preference for the upward and downward states of the perpendicular magnetization.

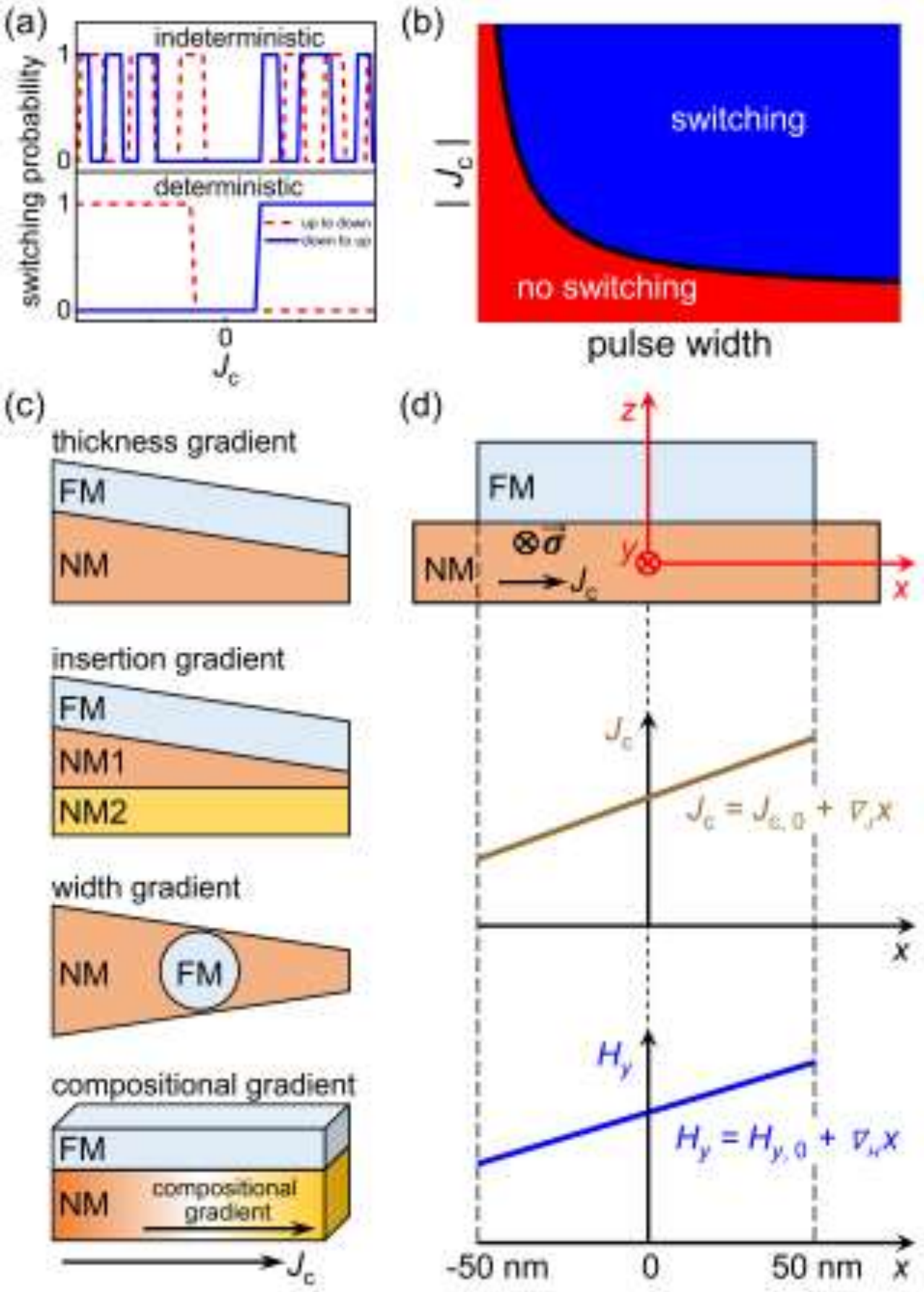

FIG. 1. (a) Schematic of switching probability as a function of pulse amplitude. (b) Phase diagram of deterministic switching. (c) Typical wedge structures for generating a longitudinal spin current density. (d) The reduced model and definitions of the parameters of the current-density gradient and the associated transverse field in simulations.

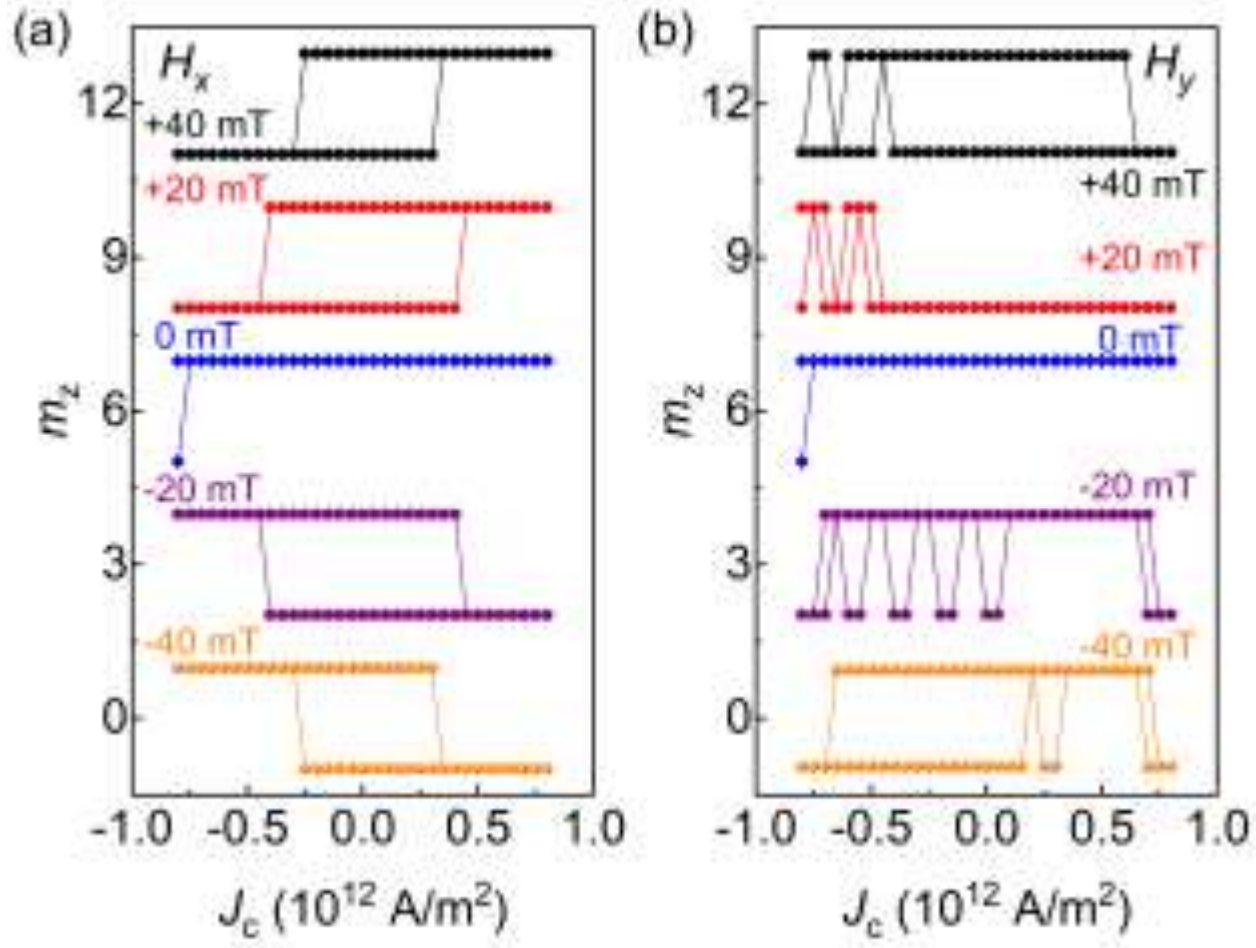

FIG. 2. Out-of-plane component of the magnetization vs the uniform current density under different uniform (a) longitudinal magnetic fields $H_x$ and (b) transverse magnetic fields $H_y$.

We further plot in Figs. 3(a) and 3(b) the time-revolved evolution of the spatially averaged out-of-plane component of the magnetization ($m_z$) during the SOT switching with different current pulse widths, without and with the current-pulse-induced transverse magnetic field, respectively. The magnetization exhibits large oscillations, while the final state depends sensitively on the pulse width, i.e., the difference between the time at which the current pulse is turned on and off. In Figs. 3(c) and 3(d), we summarize the switching results of an initially upward state device as a function of the current density $J_c$ and the pulse width ($\tau$). The device does not reverse at all until the current density reaches a very large value of 0.8×10$^{12}$ A/m$^2$. At greater current densities, the device exhibits indeterministic switching only in narrow strip-like parameter windows of tens of picoseconds in pulse width. The switching windows are slightly reshaped after the introduction of a large transverse magnetic field of $H_y$ = 10 mT [Fig. 3(d)]. Such narrow parameter windows suggest that the device can be switched only if the density, pulse width, and the transverse magnetic field of the write current pulse are very precisely constrained within the specific combinations of high $J_c$ and large $\tau$. This is technologically undesirable since it prevents reliable, stable switching required for large-scale memory and logic circuit applications in the presence of finite time-dependent drifts of the write current pulse profile.

*Effect of a current-density gradient.* In Figs. 4(a)-4(c), we plot the time-resolved evolution of $m_z$ under a 5 ns pulse current with a large density of $J_{c,0}$ = 4 ×10$^{11}$ A/m$^2$. In the absence of any transverse field ($H_y = \nabla_H \equiv 0$, Fig. 4(a)), the magnetization exhibits enhanced oscillation during the application of the current pulse due to the current gradients ($\nabla_J$ = 2 ×10$^{19}$ A/m$^3$ and 4 ×10$^{19}$ A/m$^3$) and finally a switching for $\nabla_J$ = 2 ×10$^{19}$ A/m$^3$ but not for the greater gradient $\nabla_J$ = 4 ×10$^{19}$ A/m$^3$. In the presence of a uniform (Fig. 4(b)) or gradient transverse magnetic field (Fig. 4(c)), the current with a significant gradient excites large-amplitude magnetization precession over the polar angles of >90° and, in some cases, further evolves into magnetization reversal (for $\nabla_J$ = 4 ×10$^{19}$ A/m$^3$ but not for $\nabla_J$ = 2×10$^{19}$ A/m$^3$). As the current gradient $\nabla_J$ is varied, the phase, amplitude, and relaxation of the magnetization precession oscillation are all modified. Compared to the zero transverse field case, the evolution of $m_z$ exhibits small differences that accumulate during the subsequent magnetization dynamics and ultimately lead to different final states. However, the gradient does not generate deterministic switching plotted in Fig. 1(a). In Figs. 4(d)-4(f) we plot the final switching state as a function of $J_{c,0}$ and $\nabla_J$ for a fixed pulse width of 5 ns under zero, uniform ($H_y$ = 10 mT), and gradient ($H_{y,0}$ =10 mT, $\nabla_H$ = 0.1 mT/nm) transverse field conditions, respectively. There are indeterministic switching windows only for the specific combinations of $J_{c,0}$ and $\nabla_J$. For a certain $J_{c,0}$ ($\nabla_J$), an increase in $\nabla_J$ ($J_{c,0}$) does not necessarily promote the switching event.

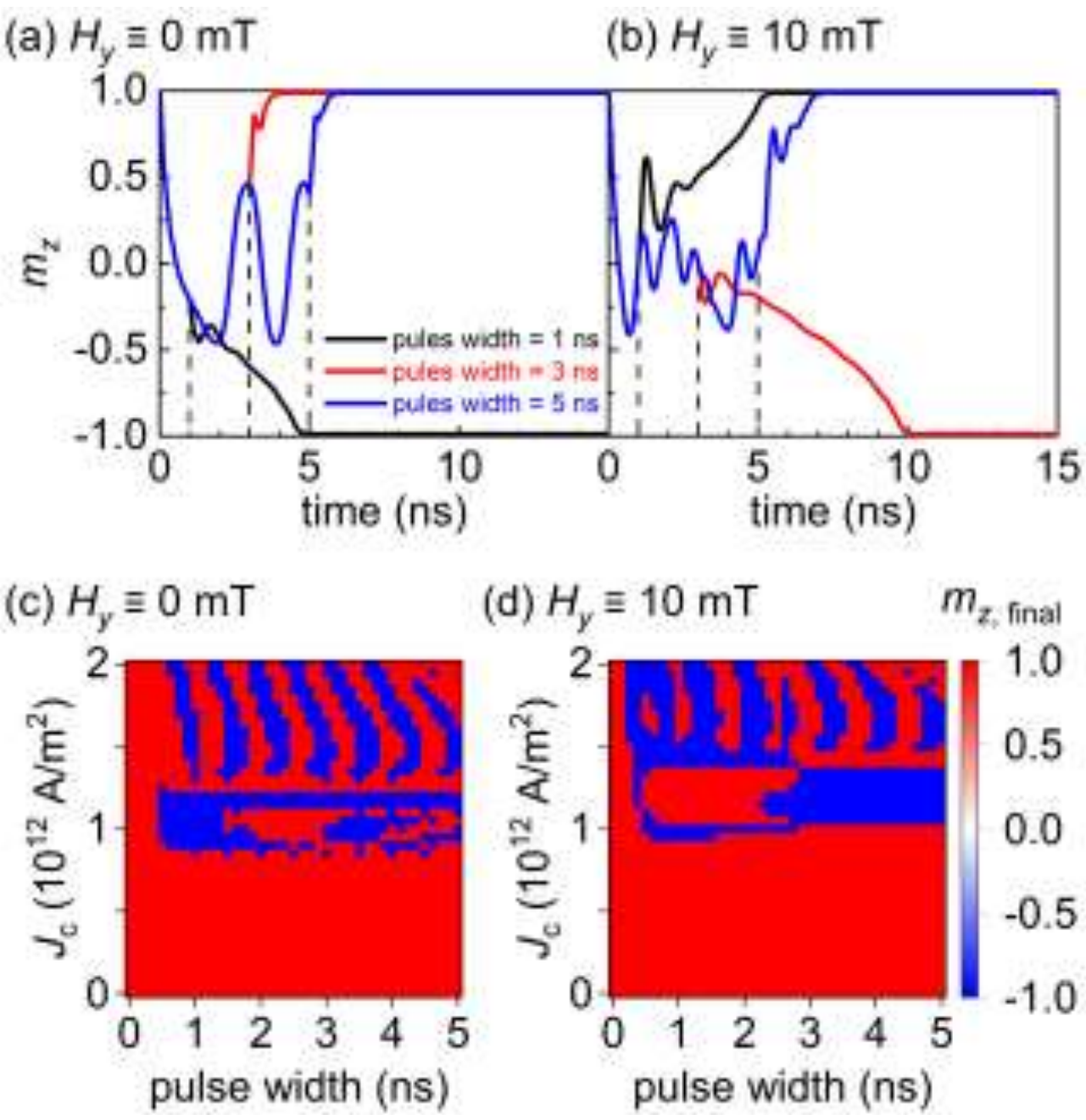

FIG. 3. Magnetization switching induced by a uniform spin current density. Time-resolved evolution of the spatially averaged out-of-plane magnetization ($m_z$) under a uniform current ($J_c = 9 \times 10^{11}$ A/m$^2$) and different pulse widths (1 ns, 3 ns, and 5 ns) simulated with the transverse magnetic field of (a) 0 mT and (b) 10 mT, respectively. Switching phase diagrams as functions of the uniform current density and pulse width with the transverse magnetic field of (c) 0 mT and (d) 10 mT. For (a)-(d), the initial state is $m_z$ = +1.

We further investigate the switching by varying $J_{c,0}$ and the pulse width by fixing a large current gradient of $\nabla_J = 2 \times 10^{19}$ A/m$^3$ in Figs. 4(g)-4(i). The switching occurs only in specific parameter windows. The separated switching windows reflect the matching of the magnetization precession phase and the pulse duration. Notably, the switching windows are also extended to the vanishingly small $J_{c,0}$ values, which differs from the uniform-current case where reversal occurs only after the current density exceeds a certain threshold [Figs. 3(c) and 3(d)]. Note that despite the small $J_{c,0}$, the large $\nabla_J$ still yields very large current density in some regions of the device (the maximum current density is always greater than $10^{12}$ A/m$^2$ for $\nabla_J = 2 \times 10^{19}$ A/m$^3$). The region with the very high current densities is switched first and further triggers the domain wall expansion towards the entire device. Comparing the cases with zero, uniform, and gradient transverse field, the switching phase windows are essentially similar, with only slight shifts. The Oersted field induced by the gradient current density can only slightly affect the transient dynamics. However, the switched parameter spaces remain isolated and window-like.

To better understand the unreliable switching behaviors, we show two representative switching processes in Figs. 5(a) and 5(b). For the unsuccessful switching event in Fig. 5(a), the device underwent complex large-portion partial switching via anti-domain nucleation and domain wall propagation due to the 5 ns current pulse ($J_{c,0} = 4 \times 10^{11}$ A/m$^2$, $\nabla_J = 2 \times 10^{19}$ A/m$^3$) but finally fell back to the initial state. During the successful switching event in Fig. 5(b), the very high current density gradient of $\nabla_J = 4 \times 10^{19}$ A/m$^3$ induces strong spatial non-uniformity in the micromagnetic dynamics, including multi-spot anti-domain nucleation and domain wall propagation. The oscillation of the averaged out-of-plane component of the magnetization in Figs. 4(a)-4(c) is mainly due to the back-and-forth motion of domain walls and the repeated nucleation and annihilation of the anti-domains. The magnetic texture contains differently oriented domains, whose subsequent evolution depends on the pulse width and the instantaneous domain configuration. After the current pulse is terminated, the strong perpendicular anisotropy of the magnetic layer drives the system to relax toward either the reversed ($m_z$ = -1) or non-reversed ($m_z$ = +1) state, depending on the instantaneous domain configuration at the end of the pulse.

It appears infeasible to predict any final states until the end of the evolution of such complex combined precession and domain wall propagation. The final state is sensitive to the finite micromagnetic dynamics tuned by the widths, amplitudes, and gradients of the current and transverse field (the sum of the Oersted field and fieldlike torque). The switching results cannot be predicted based on any simple models such as coherent macrospin rotation, deterministic domain wall depinning, or any symmetry analyses.

Therefore, a current-density gradient makes the SOT and the switching spatially inhomogeneous by influencing where the anti-domain nucleates and how the micromagnetic dynamics evolves. A longitudinal gradient in the transverse magnetic field ($H_y$, from the Oersted field and fieldlike torque) can further reshape the transient domain landscape. However, neither the longitudinal gradients in the longitudinal spin current and the transverse field can be considered an effective replacement of a longitudinal magnetic field to enable deterministic all-electrical switching.

To reaffirm the prediction of the microscopic simulations, we perform current switching experiments with Kerr microscopy. For a $Pt_{75}Ti_{25}$/Ti/FeCoB device with a strong longitudinal width gradient (the width is varied from 2 μm to 7 μm within a length scale of 5 μm, Fig. 5a), there is no switching after a large current pulse of ±4.5 mA along the giant relative current gradient of 1/μm (Fig. 5b). We note that a realistic thickness/composition wedge can only induce negligible relative charge and spin current gradients ($\leqslant 4\times10^{-5}$/μm) in practical samples. For a $Pt_{1-x}Cu_x$ with a composition wedge of $x$ = 0-1 achieved within a 4-inch wafer (which is already technically very challenging), the spin Hall angle gradient can be estimated to be less than $2\times10^{-6}$/μm, which can only generate a negligible spin current density gradient within a micro-sized Hall bar device. Thus, one can typically expect only negligible spin current density in devices with a thickness, insertion, or composition gradient in any orientation.

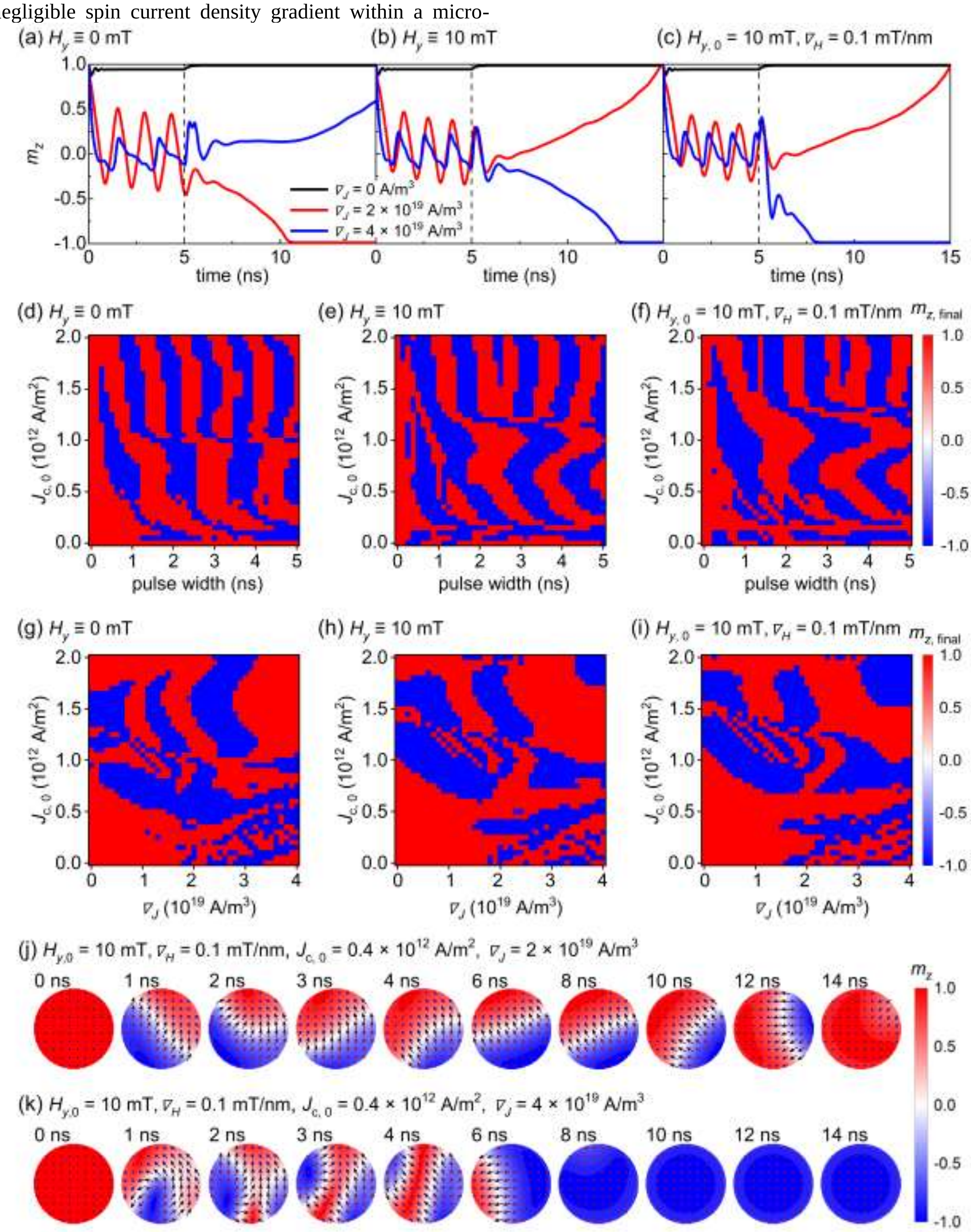

FIG. 4. Magnetization switching by a gradient spin current density. Time-resolved evolution of $m_z$ under different current-density gradients for $J_{c,0} = 4\times10^{11}$ A/m² and 5 ns pulse width under a transverse magnetic field of (a) $H_y \equiv 0$ mT, (b) $H_y \equiv 10$ mT, and (c) $H_{y,0} = 10$ mT, $\nabla_H \equiv 0.1$ mT/nm. Dependence of the switching results on the average current density and pulse width ($\nabla_J = 2\times10^{19}$ A/m³) under a transverse magnetic field of (d) $H_y \equiv 0$ mT, (e) $H_y \equiv 10$ mT, and (f) $H_{y,0} = 10$ mT, $\nabla_H \equiv 0.1$ mT/nm. Dependence of the switching results on the average current density and current gradient (pulse width = 5 ns) under a transverse magnetic field of (g) $H_y \equiv 0$ mT, (h) $H_y \equiv 10$ mT, and (i) $H_{y,0} = 10$ mT, $\nabla_H = 0.1$ mT/nm. The initial state for (a)-(i) is $m_z = +1$. Micromagnetic evolutions driven by a gradient current of (j) $\nabla_J = 2\times10^{19}$ A/m³ and (k) $\nabla_J = 4\times10^{19}$ A/m³ and a transverse magnetic field ($H_{y,0} = 10$ mT, $\nabla_H = 0.1$ mT/nm, $J_{c,0} = 0.4\times10^{19}$ A/m³).

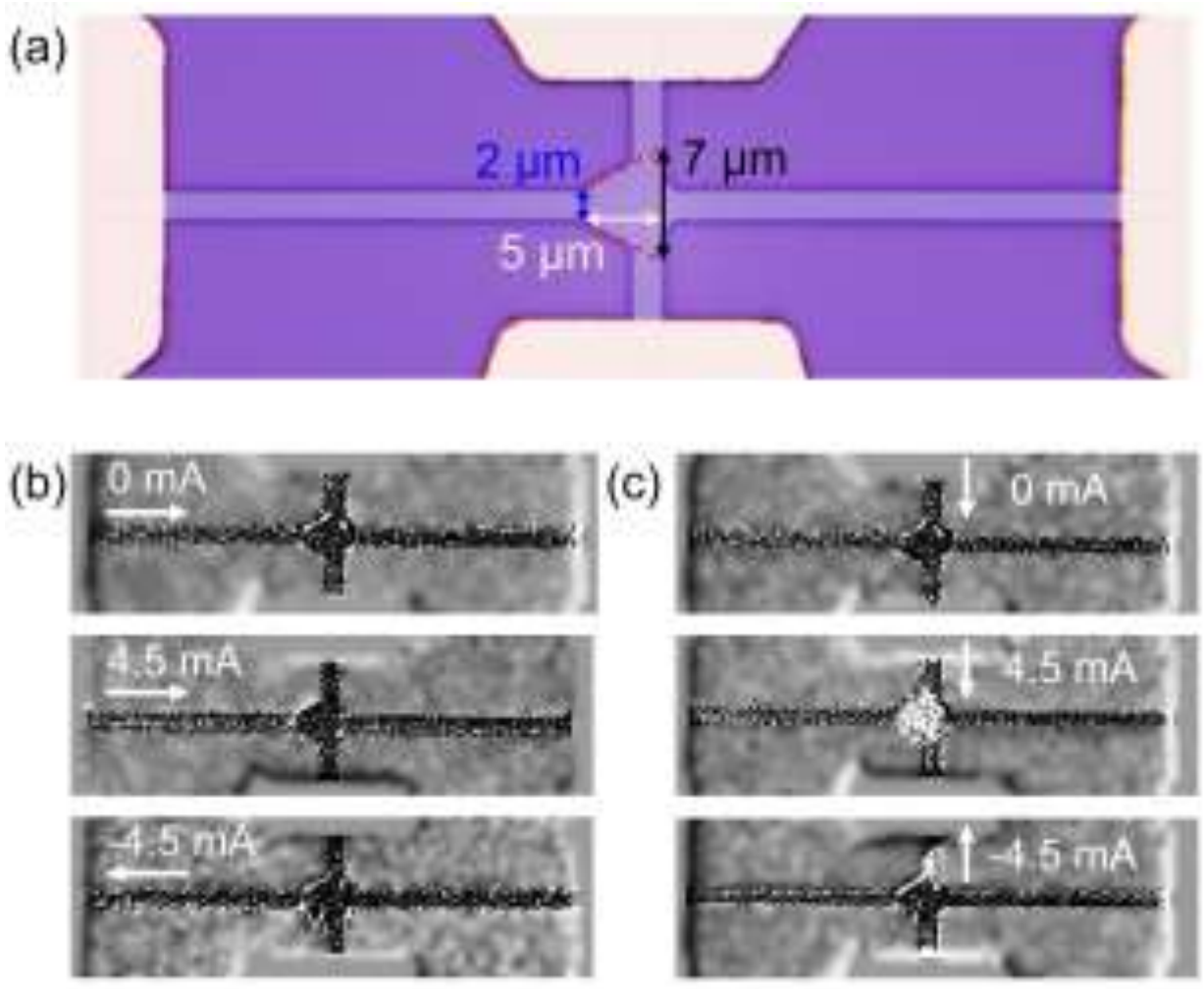

Fig. 5. Experimental verification of a $Pt_{75}Ti_{25}$/Ti/FeCoB device. (a) Optical microscopy image of the device, indicating a strong longitudinal width gradient. Kerr microscopy images after different current pulses (b) along and (c) transverse to the width gradient.

We note that, despite the insufficiency of the spin current density gradient, there could be other effective effects (e.g., $z$ spin generation) taking place in the structures with the transverse thickness/phase gradient in the NM layers [25,26] or the perpendicular composition gradients in the magnetic layers [27-32]. As shown in Fig. 5c, the same device that cannot be switched by a current along the channel width gradient exhibits deterministic switching driven by the current pulses that are of the same magnitude of $\pm 4.5$ mA but transverse to the current gradient. In this case, the electric asymmetries both *transverse* and *perpendicular* to the current (due to distinct thicknesses of the spin Hall metal layer and the electrical contacts) lead to efficient generation of perpendicular spins [33-36] and deterministically switch perpendicular magnetization via anti-damping torque, without involving any longitudinal structural, current, or spin gradients. In contrast, a longitudinal spin current density gradient, even combined with a perpendicular electric asymmetry, cannot yield generation of perpendicular spins in NM/FM heterostructures according to symmetry analysis [37,38].

*Conclusion.* We have demonstrated a combined micromagnetic and experimental study on the effects of the longitudinal spin current density gradient. We have established that any realistic longitudinal spin-current density gradient cannot be an effective replacement for a longitudinal magnetic field for deterministic switching of a PMA device. There can be only indeterministic switching windows that are isolated and picosecond-scale narrow. Technologically, such narrow indeterministic switching windows would create considerable write error rates in practical devices in the presence of finite drifts of material parameters and write current pulse profiles. Instead of a longitudinal spin current density gradient, perpendicular spins are experimentally proven to be a highly effective scheme for deterministic switching of perpendicular spin torque devices. These results suggest that the previously reported switching in magnetic heterostructures with a longitudinal gradient in composition, thickness, insertion, or channel width [14-16] was triggered by some mechanism other than spin current density gradients.

This work was supported partly by the National Key Research and Development Program of China (2022YFA1204000), the Beijing Natural Science Foundation (Grant No. Z230006) and by the National Natural Science Foundation of China (Grant No. 12274405).

**Data availability.** The data are available from the authors upon reasonable request.